\documentclass[12pt,a4paper ]{article}

\newcommand{\frat}[2]{\frac{\textstyle #1}{\textstyle #2}}
\newcommand{\vf}[1]{\mbox{\boldmath $#1$}}

\begin{document}
\begin{center}
{\Large \bf Point-like sources of Euclidean non-abelian field in
the instanton liquid}\\ \vspace{0.5cm} S.V. Molodtsov, G.M.
Zinovjev$^\dagger$\\ \vspace{0.5cm} {\small\it State Research
Center, Institute of Theoretical and Experimental Physics, 117259,
Moscow, RUSSIA}\\ $^\dagger$ {\small \it Bogolyubov Institute for
Theoretical Physics,\\ National Academy of Sciences of Ukraine,
UA-03143, Kiev, UKRAINE}
\end{center}
\vspace{0.5cm}

\begin{center}
\begin{tabular}{p{16cm}}
{\small{An appraisal of average energy of Euclidean non-abelian
point-like source in the instanton liquid is calculated. It
behaves linearly increasing at large distances while unscreened.
For the dipole in colour singlet state this energy increases
linearly again with the separation increasing and
its "tension" coefficient develops the magnitude pretty similar to
the lattice approach and other model estimates. The situation of
sources arbitrary oriented in colour space is perturbatively
considered.}}
\end{tabular}
\end{center}
\vspace{0.5cm}

Since the discovery of (anti-)instantons
(the classical solutions of the Yang-Mills
equations with nontrivial topological features)
the problem of the gauge interaction between
instantons is under intensive study and the dipole
interaction of pseudo-particles with an external field has
already been argued in the pioneering papers \cite{1}. This practically
important conclusion was grounded on using the
would-be superposition ansatz for the
approximate solution of the Yang-Mills equations
\begin{equation}
\label{1}
{\cal A}^a_{\mu}(x)=A^a_{\mu}(x)+B^a_{\mu}(x)~.
\end{equation}
Here the first term denotes the field of a single
(anti-)instanton in the singular gauge as we are planning
to consider an (anti-)instanton ensemble and need to have the
non-trivial topology at the singularity point
\begin{equation}
\label{inst}
A^a_{\mu}(x)=\frat2g~\omega^{ab}\bar\eta_{b\mu\nu}~
\frat{\rho^2}{y^2+\rho^2}~\frat{y_\nu}{y^2}~,
\end{equation}
where $\rho$ is an arbitrary parameter characterizing the
instanton size centered at the coordinate $z$ and colour
orientation defined by the matrix $\omega$, $y=x-z$,
$\bar\eta_{b\mu\nu}$ is the 't Hooft symbol (for anti-instanton
$\bar\eta \to \eta$) and $g$ is the coupling constant of
non-abelian field. The second term describes an external field.
For the sake of simplicity we consider, at first, external field
originated by immovable Euclidean colour point-like source
$e\delta^{3a}$ and Euclidean colour dipole $\pm e\delta^{3a}$ and
limit ourselves dealing with $SU(2)$ group only. With this
"set-up" we could claim now the results obtained before have
maintained the interacting terms proportional $e/g$ only and in
this paper we are aimed to analyze the contributions proportional
$e^2$ which display rather indicative behaviour of asymptotic
energy of Euclidean non-abelian point-like sources while in the
instanton liquid (IL) \cite{2}. These Euclidean sources will
generate the fields of the same nature as ones we face taking into
account gluon field quantum fluctuations $a_{qu}$ around the
classical instanton solutions $A=A_{inst}+a_{qu}$.

In the Coulomb gauge the potentials take the following forms
$$B^a_{\mu}(x)=({\vf 0},\delta^{a3}~\varphi),~\varphi=
\frat{e}{4\pi}~\frat{1}{|{\vf x}-{\vf z}_e|}~, $$
$$B^a_{\mu}(x)=({\vf
0},\delta^{a3}~\varphi),~\varphi=\frat{e}{4\pi}
~\left(\frat{1}{|{\vf x}-{\vf z}_1|}-\frat{1}{|{\vf x}-{\vf
z}_2|}\right)~, $$ for Euclidean point-like source and dipole in
colourless state, respectively, with ${\vf z}_e$ being a
coordinate of point-like source in peace and ${\vf z}_1$, ${\vf
z}_2$ as coordinates characterizing a dipole. As known the field
originated by particle source in 4d-space develops the cone-like
shape with the edge being situated just at the particle
creation. Then getting away from that point the field becomes well
established in the area neighboring to it and might be given by
the Coulomb solution. In distant area where the field penetrates
into the vacuum it should be described by the retarded solution.
Here we are interested in studying the interactions in the
neighboring field area and the cylinder-symmetrical field might be
its relevant image in 4d-space. We treat the potentials in their
Euclidean forms and then the following changes of the field and
Euclidean point-like source variables are valid $B_0\to i B_4$,
$e\to -i e$ at transition from the Minkowski space. Actually, last
variable change is resulted from the corresponding transformations
of spinor fields $\psi\to\hat\psi$,
$\bar\psi\to-i\hat\psi^\dagger$, $\gamma_0\to\gamma_4$. It means
we are in full accordance with electrodynamics where the practical
way to have a pithy theory in Euclidean space is to make a
transition to an imaginary charge.

Thus, the field strength tensor is given as
\begin{equation}
\label{2}
G_{\mu\nu}^a=\partial_\mu {\cal A}^a_{\nu}-\partial_\nu
{\cal A}^a_{\mu}+g~\varepsilon^{abc}{\cal A}^b_{\mu}{\cal A}^c_{\nu}~,
\end{equation}
where $\varepsilon^{abc}$ is the entirely asymmetric tensor and for the
field superposition of Eq.(\ref{1}) it can be given by
\begin{eqnarray}
\label{3}
&&G_{\mu\nu}^a=G_{\mu\nu}^a(B)+G_{\mu\nu}^a(A)+G_{\mu\nu}^a(A,B)~,\nonumber\\
[-.2cm]
\\[-.25cm]
&&G_{\mu\nu}^a(A,B)=g~\varepsilon^{abc}
(B^b_{\mu}A^{c}_\nu+A^b_{\mu}B^{c}_\nu )~,\nonumber
\end{eqnarray}
if $G_{\mu\nu}^a(A)$ and $G_{\mu\nu}^a(B)$ are defined as in
Eq.(\ref{2}). Then gluon field strength tensor squared reads
\begin{eqnarray}
\label{4} G_{\mu\nu}^a~ G_{\mu\nu}^a &=&
G_{\mu\nu}^a(B)~G_{\mu\nu}^a(B) + G_{\mu\nu}^a(A)~G_{\mu\nu}^a(A) +
G_{\mu\nu}^a(A,B)~G_{\mu\nu}^a(A,B) +\nonumber\\ [-.2cm]
\\[-.25cm]
&+&2~ G_{\mu\nu}^a(B)~G_{\mu\nu}^a(A) + 2~
G_{\mu\nu}^a(B)~G_{\mu\nu}^a(A,B) + 2~
G_{\mu\nu}^a(A)~G_{\mu\nu}^a(A,B)~. \nonumber
\end{eqnarray}
The various terms of Eq.(\ref{4}) provide the contributions of
different kinds to the total action of full initial system of
point-like sources and pseudo-particle
\begin{equation}
\label{5}
S=\int dx \left( \frat14~ G_{\mu\nu}^a~G_{\mu\nu}^a+j^a_{\mu}
{\cal A}^a_{\mu}\right)~.
\end{equation}
So, the first and second terms of Eq.(\ref{4}) provide the source
self-energy (for the dipole it should be supplemented by the
Coulomb potential of interacting sources) and the single instanton
action $\beta=\frat{8\pi^2}{g^2}$. At small distances the first term
is regularized by introducing the source "size" as in classical
electrodynamics. These terms are proportional to $e^2$ and $g^{-2}$,
respectively. The fourth and last terms of Eq.(\ref{4}) together with the
term $j^a_{\mu}A^a_{\mu}$ of Eq.(\ref{5}) provide the contribution of
proportional $e/g$ and the third term of Eq.(\ref{4}) only leads to the
contribution proportional $e^2$ as the fifth term is equal zero
owing to the gauge choice. Denoting the non-zero contributions as
$S_{int}$ we have after performing the calculations
\begin{equation}
\label{6}
S_{int}=\frat{e}{g}~\bar\eta_{k4i}~\omega^{3k} I_i
+\left(\frat{e}{4\pi}\right)^2 J+ \left(\frat{e}{4\pi}\right)^2
K_{kl}~\omega^{3k}\omega^{3l}~.
\end{equation}
The exact form of $I_i$ is unnecessary in this paper as applying
the results for the IL model we have to average over the colour
orientation of (anti-)instanton what leads to disappearance of
dipole contribution. Two other terms are resulted from the "mixed"
component of field strength tensor
\begin{eqnarray}
\label{mc} &G_{4i}^a(A,B)=2~\frat{e}{4\pi}~\varepsilon^{a3c}~
\omega^{ck}~\bar\eta_{ki\alpha}~
\frat{y_\alpha}{y^2}~\frat{\rho^2}{(y^2+\rho^2)}~\frat{1}{|{\vf
y}+{\vf \Delta}|}~, &\nonumber\\[-.2cm]
\\[-.25cm]
&G_{4i}^a(A,B)=2~\frat{e}{4\pi}~\varepsilon^{a3c}~\omega^{ck}
~\bar\eta_{ki\alpha}~
\frat{y_\alpha}{y^2}~\frat{\rho^2}{(y^2+\rho^2)}~
\left(\frat{1}{|{\vf y}+{\vf \Delta}_1|}-\frat{1}{|{\vf y}+{\vf
\Delta}_2|}\right)~,&\nonumber
\end{eqnarray}
to correspond to the point-like source and to the field of
colourless dipole where
${\vf \Delta}={\vf z}-{\vf z}_e$,
${\vf \Delta}_{1,2}={\vf z}- {\vf z}_{1,2}$.
The other contributions to $G^a_{\mu\nu}(A,B)$ are absent because
of the gauge used. In order to handle the further formulae easily
it is practical to introduce new dimensionless coordinates as
$x/\rho\to x$. Then for the single source the function $J$ and the
tensor $K$ take the following forms
$$J=2~\int dy~\frat{2~y^2-
{\vf y}^2}{y^4~(y^2+1)^2~|{\vf y} + {\vf \Delta}|^2}~,$$
$$K_{kl}=2~\int dy~ \frat{y_ky_l}{y^4}~\frat{1}{(y^2+1)^2~|{\vf y}
+ {\vf \Delta}|^2}~,$$ and can not be integrated in the elementary
functions. Fortunately, we need their asymptotic values at
$\Delta\to\infty$ only $$J\simeq
\frat{5\pi^2}{2}\frat{1}{\Delta^2}~,$$ and for the components of
the 2-nd rank tensor
$$K_{ij}=\delta_{ij}~K_1+\hat\Delta_i\hat\Delta_j~K_2~,$$
we have
$$K_1\simeq\frat{\pi^2}{2}\frat{1}{\Delta^2}~,~~K_2\simeq 0~.$$

Clearly, the mixed component of field strength tensor is of purely
non-abelian origin but its contribution to the action of whole
system (point-like sources and pseudo-particle) takes the form of
self-interacting Euclidean source $\sim e^2$ although it is
constructed by the instanton field and field generated by sources.
Seems, this simple but still amazing fact was not explored
properly.

Now we are trying to analyze the pseudo-particle behaviour in the
field of Euclidean non-abelian source developing the perturbative
description related to the pseudo-particle itself. Realizing
such a program we "compel" the pseudo-particle parameters to be the
functions of "external influence", i.e. putting $\rho\to R(x,z)$,
$\omega\to\Omega(x,z)$. These new fields-parameters are calculated
within the multipole expansion and then, for example, for the
(anti-)instanton size we have
\begin{eqnarray}
\label{dec} R_{in}(x,z)&=&\rho+c_\mu~y_\mu+c_{\mu\nu}~y_\mu~
y_\nu+\dots~,~~~~~|y|\leq L \nonumber\\ [-.2cm]
\\[-.25cm]
R_{out}(x,z)&=&\rho+d_\mu~\frat{y_\mu}{y^2}+d_{\mu\nu}~
\frat{y_\mu}{y^2}~\frat{y_\nu}{y^2}+\dots~,~~~|y|>L~, \nonumber
\end{eqnarray}
Similar expressions could be written down for the (anti-)instanton
orientation in the colour space $\Omega(x,z)$ where $L$ is a
certain parameter fixing the radius of sphere where the multipole
expansion growing with the distance increasing should be changed
for the decreasing one being a result of deformation regularity
constraint imposed. Then the coefficients of multipole expansion
$c_\mu,c_{\mu\nu},\dots$ and $d_\mu,d_{\mu\nu},\dots$ are the
functions of external influence. It turns out this approach allows
us to trace the evolution of approximate solution for the deformed
(crumpled) (anti-)instanton as a function of distance and,
moreover, to suggest the selfconsistent description of
pseudo-particle and point-like fields within the superposition
ansatz of Eq.(\ref{1}) (the paper has been completed recently).

Proceeding to the calculation of average energy of point-like
source immersed into IL we have to remember that one should work
with the characteristic configuration saturating the functional
integral being the superposition of (anti-)instanton fields also
supplemented by the source field $B^a_{\mu}(x)$, i.e.
\begin{equation}
\label{8} {\cal A}^{a}_\mu(x)=B^a_{\mu}(x)+\sum_{i=1}^N
A^{a}_\mu(x;\gamma_i)~,
\end{equation}
where $\gamma_i=(\rho_i,z_i,\omega_i)$ are the parameters
describing the $i$-th (anti-)instanton. The IL density at large
distances from source coincides practically with its asymptotic
value $n(\Delta)\sim n_0 e^{\beta-S}\simeq n_0$ because an action
of any pseudo-particle there is approximately equal $\beta$. The
quantity we are interested in should be defined by averaging $S$
over the pseudo-particle positions and their colour orientations
(taking all pseudo-particles of the same size) $\prod_{i=1}^N
\frat{d z_i}{V}~ d\omega_i$, where $V$ is the IL volume, and reads
$$E=\frat{\langle \int d x~ G^a_{\mu\nu}G^a_{\mu\nu}\rangle}{X_4}=
\frat{e^2}{4\pi}\frat{1}{a}+\langle S'-\beta\rangle_3~.$$ The
first term where $a$ is a source "size" value (on strong interaction
scale, of course) corresponds to the source energy and the second
term presents contribution of all (anti-)instantons developing in
thermodynamic limit the following form
\begin{equation}
\label{s}
\langle S'-\beta \rangle_3\simeq \langle
S_{int}\rangle_3 \sim n_3 \int d {\vf \Delta}
\left(\frat{e}{4\pi}\right)^2\left(J+\frat{K_{ii}}{3}\right)~,
\end{equation}
with $n_3=n^{3/4}$ being the IL density in 3d space $(n=N/V)$ and
$S'$ means the action without contribution of the first term of
Eq.(\ref{4}). In fact, the result can be reduced to the common
denominator $X_4$ (playing the role of "time") as our concern here
is (anti-)instanton behaviour in the source background when the
field is well developed and the solution possesses an automodel
property at any $x_4$ layer. The constant originated by the second
term of Eq.(\ref{4}) was omitted.

Then returning to the dimensional variables for the moment and
using the asymptotic behaviours of corresponding magnitudes one
can easily examine a linearly increasing contribution of source
self-energy type
$$\langle S'-\beta \rangle_3\sim \frat{6\pi^3}{\beta} \left(\frat{\bar\rho}{\bar
R}\right)^3\frat{1}{\bar\rho}\frat{L}{\bar\rho}=\sigma~L~,$$
where
$L$ is a formal upper limit of integration, $\bar\rho$ is the mean
(anti-)instanton size , $\bar R$ denotes the pseudo-particles
separation and it looks natural for the source to have $e=g$. For
the $SU(N_c)$-group the denominator of Eq.(\ref{s}) should be
rearranged according to the change $3\to N_c^{2}-1$ but there is
no any noticeable impact on the result obtained. The "tension"
coefficient at the parameters $\frat{\bar\rho}{\bar R}\simeq
\frat{1}{3.67}$, $\beta\simeq 18$, $\bar\rho\simeq (1$ GeV$)^{-1}$
being characteristic for IL \cite{3} takes the value of $\sigma
\simeq 1$ GeV/fm what is in full accordance with the estimates
extracted from the models for heavy quarkonia, for example. If one
intends to explore the magnitudes like $\langle
S_{int}~e^{-S_{int}}\rangle$ (which could model an effect
of suppressing pseudo-particle contribution in the source vicinity)
in numerical calculations it becomes evident the linearly
increasing behaviour starts to form at $\Delta/\bar\rho\sim
3$---$4$. The well known fact of area law absent for the
Wilson loops in the pure gluodynamics could be
explained by rapid decay ($\sim \Delta^{-4}$) of the
corresponding correlators in IL and inadequate contribution of
large size instantons. Apparently, for present consideration the
result is rooted in non-abelian nature of gauge field and treating
the approximate solution in the form of superposition ansatz of
pseudo-particle and source fields. Perhaps, this picture could be
made physically more transparent if other observables, for
example, the non-abelian field flux over a surface are considered.
However, we believe the result already obtained delivers one
interesting message. The energy growth with the distance
increasing teaches about impossibility of bringing the Euclidean
colour source in IL as the source mass (the additional
contribution received should be treated just in this way) is
unboundedly increasing if the screening effects do not enter the
play. This asymptotic estimate of Euclidean source energy in IL
provides the major contribution to the generating functional in
quasi-classical approximation if all the coupling constants are
frozen at the scale of mean instanton size $\bar\rho$.

Now dealing with the field of colour dipole in the colour singlet
state we are able to demonstrate the IL reaction once more. The
contribution to average energy $\langle S'-\beta \rangle_3$ which
we are interested in is defined at large distances by the
integral very similar to that for the configuration with one
single source. It has even the same coefficient but another averaging over
the (anti-)instanton positions as
$$I_d=\int d{\vf \Delta}_1 \left(\frat{1}{{\vf\Delta}_1^{2}} -
2~\frat{1}{|{\vf\Delta}_1| |{\vf\Delta}_2|}+
\frat{1}{{\vf\Delta}_2^{2}}\right)~.$$
When the source separation $l=|{\vf z}_1-{\vf z}_2|$ is going to zero, the
field disappears and the final result should be zero. There are
two parameters only to operate with the integral, they are $L$
and $l$. The dimensional analysis teaches the integral is the
linear function of both but the $l$-dependence only obeys the requirement
of integral disappearance at $l\to 0$. It is easy to receive the
equation
$$\frat{I_d}{4\pi}=L-2\left(L-\frat{l}{2}\right)+L=l~$$
for determining the coefficient (the contributions of three
integrals are shown separately here). Finally the average dipole
energy reads $$\langle S'-\beta\rangle_3\sim \sigma~l~.$$

Recently it becomes well known that for the SU(2)-group two
point-like sources problem can be resolved for arbitrary
orientation of the sources in colour space [4,5].
Moreover, the
self-consistent scheme to describe the source (particle) and
fields dynamics in the form of $v/c$-expansion can be elaborated
as in electrodynamics \cite{5}. Now in order to make the further
conclusions understandable we remember some necessary results. Let
us suppose in the points ${\vf z}_{1,2}$ the sources of intensity
$e \widetilde P_1$, $e \widetilde P_2$ are situated,
correspondingly, where the tilde sign on the top $\widetilde
P=(P^1,P^2,P^3)$ means the vector in colour space with the unit
normalization $|\widetilde P|=(P^\alpha P^\alpha)^{1/2}=1$. These
source vectors can be treated as a convenient basis to span the
Yang-Mills solutions
\begin{eqnarray}
\label{a1}&& \widetilde B_4 =\varphi_1 \widetilde P_1+\varphi_2
\widetilde P_2~, \nonumber\\ [-.2cm]
\\[-.25cm]
&&\widetilde {\vf B} ={\vf a}~ \widetilde P_1 \times \widetilde P_2
~.\nonumber
\end{eqnarray}
If one requires now the fields are going to disappear at large
distances from sources then the basis of vector-sources rotating
around the constant vector $\widetilde \Omega
=\stackrel{*}{\varphi}_1 \widetilde P_1+ \stackrel{*}{\varphi}_2
\widetilde P_2$ could correspond to such a choice of gauge
\begin{eqnarray}
\label{a2} \dot{\widetilde P}_1=g~ \widetilde B_4({\vf z}_1)
\times \widetilde P_1~, \nonumber\\ [-.2cm]\\[-.25cm]
\dot{\widetilde P}_2=g~ \widetilde B_4({\vf z}_2) \times
\widetilde P_2~, \nonumber
\end{eqnarray}
with the frequency $g|\widetilde \Omega |$ where
$\stackrel{*}{\varphi_1}=\varphi_1({\vf z}_2)$,
$\stackrel{*}{\varphi_2}=\varphi_2({\vf z}_1)$. The vectors dotted on their
top mean the differentiation in $x_4$. The same character of solutions persist
at transition to the Minkowski space. The functions $\varphi_{1,2}$ and
vector-function ${\vf a}$ are dependent on ${\vf x}$
only and are determined by the following equations
\begin{eqnarray}
\label{a3}
{\vf D}{\vf D}( \varphi-\stackrel{*}{ \varphi})=-e{ \delta} \nonumber\\ [-.2cm]
\\[-.25cm]
{\vf \nabla} \times {\vf \nabla} \times {\vf a}={\vf j}, \nonumber
\end{eqnarray}
where ${\vf D}_{kl}=\nabla \delta_{kl}+g{\vf a} C_{kl},~ k,l=1,2$
and the current is defined as
$${\vf j}=g~( \varphi-\stackrel{*}{
\varphi}) J {\vf D} ( \varphi-\stackrel{*}{ \varphi})~.$$ Besides,
the operator ${\vf D}$ is also acting on $$ \varphi^{\mbox{\tiny
T}}=\| \varphi_1,\varphi_2 \|~,~~ \stackrel{*}{
\varphi^{\mbox{\tiny T}}}=\|\stackrel{*}{\varphi_1},
\stackrel{*}{\varphi_2}\|~,~~
\delta^{\mbox{\tiny T}}=\|\delta ({\vf x}-{\vf z}_1),
\delta ({\vf x}-{\vf z}_2)\|~.\nonumber
$$
The matrices $C$ and $ J$ are defined as

\vspace{0.5cm} \parbox[b]{3.6in}{$ C=
\left \| \begin{array}{rr}
-(\widetilde P_1 \widetilde P_2) & -(\widetilde P_2 \widetilde P_2) \\
(\widetilde P_1 \widetilde P_1) & (\widetilde P_1 \widetilde P_2)
\end{array} \right \|~,$}
\parbox[b]{3.6in}{$
J=
\left \| \begin{array}{cc}
0 & 1 \\
-1 & 0
\end{array} \right \|~.$}
\vspace{0.3cm}

It is resulted from Eq.(\ref{a2}) the modules of vectors
$\widetilde P_{1,2}$ and their scalar product $(\widetilde
P_1\widetilde P_2)$ do not change in "time". The system of
Eq.(\ref{a3}) has pretty transparent physical meaning. The colour
field originated by two point-like sources is an origin of colour
charge itself because the gluons are not neutral. The
self-consistent picture of charges and corresponding currents is
set up in between the initial charges. The solutions of this
system were accurately investigated both analytically and
numerically and it has been found out the sources are interacting
in the Coulomb-like way at any magnitude of coupling constant $g$
and when the coupling constant is not large $\frat{g}{4\pi}<
\sqrt{2}$, the solutions are well approximated by simple
Coulomb-like potentials
$$\varphi_{1,2}=\frat{e}{4\pi}~\frat{1}{|{\vf x}-{\vf
z}_{1,2}|}~$$ with ${\vf a}$ defined by the current density
generated by these potentials. In general, the vector field looks
like the field of constant magnet with the poles placed in the
points ${\vf z}_{1},{\vf z}_{2}$. Then at the straight line
connecting sources the field develops the longitudinal component
of constant value only $$|{\vf
a}_{\|}|=\frat{e}{4\pi}~\frat{eg}{|{\vf z}_1-{\vf z}_{2}|}~.$$
Going away from this straight line the field is rapidly getting
weaker. The total energy stored in the colour field is evaluated
as $$E_{cf}=\int d {\vf x}~ \frat{\widetilde {\vf E}^2+\widetilde
{\vf H}^2}{2}\sim e^2~\frat{(\widetilde P_1\widetilde P_2)}{l}+g^2
e^4~I~\frat{ (\widetilde P_1\times\widetilde P_2)^2}{l}~ $$ where
$I=\pi~(6-\pi^2/2)$. Apparently, we did not include the terms
corresponding to the self-interaction of sources. When the source
separation becomes large enough the frequency of precession is
swiftly going to zero $\sim 1/l$, and the basis composed by the
vectors $\widetilde P_1,\widetilde P_2, \widetilde
P_1\times\widetilde P_2$ could be approximated as being in peace.
The contribution of nonsingular vector field (comparing to the
Coulomb field, see Eq.(\ref{mc})) to the interaction with
(anti-)instanton may be neglected. Then the simple superposition
$$\widetilde B_4 =\varphi_1 \widetilde P_1+\varphi_2 \widetilde
P_2~$$ might be used as an approximate solution. Clearly, the
contribution of such a dipole to the mean energy of IL at large
distances is proportional to $$\frat{I_d}{4\pi}=(\widetilde P_1
\widetilde P_1)L+2~( \widetilde P_1 \widetilde
P_2)\left(L-\frat{l}{2}\right)+ (\widetilde P_2 \widetilde P_2)L=
(\widetilde P_1+ \widetilde P_2)^2~L- ( \widetilde P_1 \widetilde
P_2)~l~.$$ At small distances where the Coulomb-like fields are
large the (anti-)instantons are strongly suppressed and this
factor together with the contribution of vector field ${\vf a}$
should be taken into account. Here we limit ourselves with this
estimate obtained only. Surely, the result demonstrates again that
it is very difficult to reveal the states with open colour in IL
and there is the small parameter in the problem. Indeed, the
deviations of vector-sources from the anti-parallel orientations
may not be large since it is reasonable to estimate $L\sim R_D$
screening radius for which in the approach developed $R_D\ge
m_p/\sigma$ is valid where $m_p$ is the mass of lightest stable
(on the scale of strong interaction) particle of non-Goldstone
nature.

Thus, we described all field states at any positions of sources
${\vf z}_1$, ${\vf z}_2$ for any source orientations $\widetilde
P_1$, $\widetilde P_2$. The estimate of these configurations
contribution to the generating functional can be performed within
the variational maximum principle. Then the calculated mean energy
of Euclidean sources appears in the exponential factor and with a
precision up to several inessential terms may be presented as the
suppression factor for the states with open colour
$$Z \ge e^{-E X_4 }~,$$
we omitted the condensate contribution and the part of
fermion component here. The sources could be interpreted as
non-relativistic particles if they have large masses $m_{1,2}$ and
their coordinates are the functions of "time" $x_4$ similar to
their states in colour space where they are described by the
following spinors $u^T=(1,0)$ and $\bar u^T=(0,1)$. Assuming the
changes of these states to be insignificant we describe their
supposed evolution perturbatively with the matrices $U\simeq
1+i~{\vf \sigma}{\vf \lambda}/2$, $V\simeq 1+i~{\vf \sigma}{\vf
\mu}/2$ (here $\vf\sigma$--- are the Pauli matrices) for the first
and second particles, respectively. As the result we have for the
vector-source of first particle $P^a_{1}\simeq
\delta^{3a}-\varepsilon^{3ab} \lambda^b$ and similarly for the
second one $P_2^{a}\simeq -\delta^{3a}-\varepsilon^{3ab} \mu^b$.

Hence, if the suppression factor is interpreted according to the
potential model philosophy then the generating functional in the
form
\begin{eqnarray}
\label{gen}&& Z=\int D[{\vf \lambda}]~D[{\vf z}_1]~D[{\vf
\mu}]~D[{\vf z}_2] ~ e^{-S}~,\nonumber\\
[-.2cm]\\[-.25cm] &&S\simeq\int d x_4 \left(~ \frat{{\vf
p}^2_{1}}{2m_1}+\frat{{\vf p}^2_{2}}{2m_2}+ \sigma~( \widetilde
P_1+ \widetilde P_2)^2~R_D- \sigma~( \widetilde P_1 \widetilde
P_2)~|{\vf z}_1-{\vf z}_2|+e^2\frat{(\widetilde P_1\widetilde
P_2)} {|{\vf z}_1-{\vf z}_2|}~\right)~,\nonumber
\end{eqnarray}
will be adequate to the quantum mechanical system of coloured
particles. It describes the evolution of the states $\Psi(x_4;{\vf
z}_1,{\vf z}_2;{\vf \lambda}, {\vf \mu})$ where ${\vf p}_i$ are
the particle momenta ( the particle spins are not taken into
account). Finally, the generating functional should be presented
as an integral over "coloured" spinors. However, Eq.(\ref{gen}),
seems, underestimates the factor of "coloured" spinor evolution
because the first and second components of the vectors
${\vf\lambda}$ and ${\vf\mu}$ only are essential at integrating,
i.e. Eq.(\ref{gen}) should be taken as an approximate expression.
The Coulomb-like part of this expression looks similar to the
electrodynamics and in the limit of the IL density going to zero
the expression should properly reproduce the results for the
(anti-)parallel sources.

Nevertheless, very instructive message of Eq.(\ref{gen}) is that
the system prefers to evolve over colourless states. The
probability of getting the source orientation out of the
anti-parallel position is strongly suppressed by the factor
$\sigma R_D$. It is clear the problem admits various
generalizations (number of particles, identical particles, etc)
but apparently the result will provide again a strong hint that
the integrations over the colourless states only are essential. In
a sense, it demonstrates a sort of duality of integrating over the
"coloured" spinors and the colourless (hadronic) states.
Concluding we would like to emphasize the proposed approximation
for calculating the generating functional for sources of Euclidean
non-abelian field within quasi-classical approach leads to rather
indicative picture of the dynamics of the objects resembling many
features of strongly interacting particle phenomenology.

\noindent The authors are partly supported by STCU \#P015c,
CERN-INTAS 2000-349, NATO~2000-PST.CLG 977482 Grants.

\end{document}